\documentclass[twocolumn,aps,prb,showpacs]{revtex4}
\usepackage{epsfig}
\usepackage{amsmath}
\usepackage{amsfonts}
\usepackage{amssymb}

\begin{document}

\author{Yaroslav Tserkovnyak}
\author{Gregory A. Fiete}
\altaffiliation{Present address: Kavli Institute for Theoretical Physics, UC Santa Barbara, CA 93106, USA}
\author{Bertrand I. Halperin}
\affiliation{Lyman Laboratory of Physics, Harvard University,
 Cambridge, Massachusetts 02138, USA}

\title{Mean-field magnetization relaxation in conducting ferromagnets}

\begin{abstract}
Collective ferromagnetic motion in a conducting medium is damped
by the transfer of the magnetic moment and energy to the itinerant
carriers. We present a calculation of the corresponding
magnetization relaxation as a linear-response problem for the
carrier dynamics in the effective exchange field of the
ferromagnet. In electron systems with little intrinsic spin-orbit
interaction, a uniform magnetization motion can be formally
eliminated by going into the rotating frame of reference for the spin dynamics.
The ferromagnetic damping in this case grows linearly with the
spin-flip rate when the latter is smaller than the exchange
field and is inversely proportional to the spin-flip rate in
the opposite limit. These two regimes are analogous to the
``spin-pumping'' and the ``breathing Fermi-surface'' damping
mechanisms, respectively. In diluted ferromagnetic semiconductors,
the hole-mediated magnetization can be efficiently relaxed to the
itinerant-carrier degrees of freedom due to
the strong spin-orbit interaction in the valence bands.
\end{abstract}
\date{\today}
\maketitle


Relaxation of the ferromagnetic magnetization dynamics
is well understood phenomenologically, being often accounted
for by a single dimensionless parameter, the so-called Gilbert damping
$\alpha$.\cite{Gilbert:pr55} The equation of motion conserving the
magnetization magnitude is written for the local magnetization-direction
unit vector $\mathbf{m}$ as
\begin{equation}
\partial_t\mathbf{m}=-\gamma\mathbf{m}\times\mathbf{H}_{\text{eff}}+\alpha\mathbf{m}\times\partial_t\mathbf{m}\,,
\label{llg}
\end{equation}
where $\gamma$ is the (minus) gyromagnetic ratio.
The first term on the right-hand side
describes the motion of the magnetization, $\mathbf{M}=M_s\mathbf{m}$,
in the effective field
$\mathbf{H}_{\text{eff}}=-\partial_{\mathbf{M}}E[\mathbf{M}]$,
which preserves the magnetic energy $E[\mathbf{M}]$ in the presence of applied,
crystal, exchange, and demagnetization fields. \cite{Landau:80}
The second term characterizes the dissipation of the magnetic energy due to
coupling with other degrees of freedom.
In the case of small-angle motion near an equilibrium
rotational-symmetry axis,
Eq.~(\ref{llg}) describes a damped circular precession with frequency
$\omega=\gamma H_{\text{eff}}$. In the presence of
anisotropies, both $\omega$ and $\alpha$ become tensor
quantities, and the trajectories elliptic.
For the purpose of our discussion, it is sufficient to
treat the simple case of circular precession with a scalar damping
$\alpha$.

Despite decades of experimental and theoretical studies of itinerant
ferromagnetism in metals and, more recently, in semiconductors,
the microscopic origin of $\alpha$ is still
not fully understood. One possible proposed mechanism involves a
transfer of the angular momentum (and energy) of a nonequilibrium
ferromagnetic configuration to the itinerant electrons via the exchange
interaction, with a subsequent spin-orbit relaxation to the
lattice. Such a process has been studied extensively within
the $s-d$ model, see, e.g., Refs.~\onlinecite{Mitchell:pr57,Heinrich:pss67},
although its implied applicability to the itinerant transition-metal
ferromagnetism has not been demonstrated.
The $s-d$ picture was resurrected recently\cite{Sinova:prb04}
to address the question of magnetization relaxation in the
ferromagnetic semiconductor (Ga,Mn)As, where the
ferromagnetism originates in the hole-mediated exchange interaction
between the substitutional (paramagnetic) spin-5/2 Mn
atoms.\cite{Ohno:mmm99} This letter puts forward a
description of the magnetization damping due to the exchange interaction
between the localized magnetic orbitals and the itinerant carriers in
ferromagnetic metals and semiconductors, by reducing the
problem to treating the carrier dynamics in a time-dependent uniform
exchange field.

There is at least one Gilbert-damping mechanism which can be identified and
measured separately from the others in metallic ultrathin films.
Its origin is nonlocal: The motion of a small
ferromagnet with a large surface-to-volume ratio pumps spins into
adjacent conductors, which can then be damped by spin-flip scattering
outside of the ferromagnet, leading to the Gilbert form of relaxation
with an $\alpha$ which can dominate the intrinsic
damping\cite{Tserkovnyak:prl021}. This process was
thoroughly studied experimentally in ultrathin
films\cite{Mizukami:jjap01,Ingvarsson:prb02}
giving remarkable agreement with the parameters obtained by
first-principles band-structure calculations.\cite{Zwierzycki:cm04}
Bolstered by the success in understanding the
enhanced damping of thin films by using a picture of the ferromagnetic
relaxation via time-dependent exchange interaction with itinerant carriers,
we apply related ideas to formulate a framework for studying intrinsic
relaxation of conducting ferromagnets.

Consider an $sp-d$ model of a conducting ferromagnet, where
the spins $\mathbf{J}$ of the itinerant $s$ or $p$ orbitals (either electrons or
holes) are polarized by an exchange field
$\Omega$ along the magnetization direction $\mathbf{m}$ of localized
$d$ orbitals:\cite{noncollinearity comment}
\begin{equation}
H(t)=\mathcal{H}-\Omega\mathbf{m}(t)\cdot\mathbf{J}\,.
\label{H}
\end{equation}
Here $\mathcal{H}$ is a (time-independent) Hamiltonian which
depends on the host band structure.
The exchange field can be induced by the localized
paramagnetic impurities, such as substitutional Mn atoms in
(Ga,Mn)As. Although such exchange can be highly nonuniform
on the atomic scales, we are making a simplifying assumption in
Eq.~(\ref{H}) of a uniform field whose magnetization direction $\mathbf{m}$
is treated classically in the mean-field approximation.
In particular, the magnetization is taken to be spatially
uniform on the relevant length scales of the carrier dynamics. 
We choose as a concrete example for some of our discussions the spherical
Luttinger Hamiltonian for the spin-$3/2$ holes in the valence bands of
a dilute p-doped semiconductor (e.g., GaAs, Si, or Ge):
\begin{equation}
\mathcal{H}=(2m_e)^{-1}\left\{\left[\gamma_1+(5/2)\gamma_2\right]p^2-2\gamma_2(\mathbf{p}\cdot\mathbf{J}/\hbar)^2\right\}\,,
\label{H0}
\end{equation}
where $m_e$ is the free-electron mass and the $\gamma_i$ are the
so-called Luttinger parameters.\cite{Luttinger:pr56} The spin-orbit
term couples the hole momentum $\mathbf{p}$ with its spin
$\mathbf{J}$. For the
validity of the four-band model (\ref{H0}), the carrier density must be
low enough that the Fermi energy is smaller than the intrinsic host
spin-orbit interaction energy. For the discussion of
spin-$1/2$ electron systems, we set $\gamma_2=0$.
Suppose the magnetization of the localized orbitals varies slowly in
time (being uniform at all times),
so that the time-dependent $\mathbf{m}$ modulates the
Hamiltonian (\ref{H}) adiabatically. This means that the system
equilibrates on time scales faster than the motion of $\mathbf{m}$ and
all the quantities parameterizing the carrier Hamiltonian stay constant.
Many-magnon scattering\cite{Arias:prb99} is disregarded. Such a
time-dependent long-range ferromagnetic order can be achieved
in ferromagnetic resonance (FMR) experiments on thin films
of transition metals\cite{Heinrich:ap93} and
semiconductors.\cite{Goennenwein:apl03}

Consider the small-angle dynamics of the unit vector $\mathbf{m}(t)$ in Eq.~(\ref{H}) near the $z$ axis. Suppose the equilibrium value of the average spin is collinear with the magnetization. The variation $\delta\mathbf{m}(t)=\mathbf{m}(t)-\mathbf{\hat z}$ will induce the uniform spin density $\delta j_x(\omega)=\chi_{j_xj_x}(\omega)\Omega\delta m_x(\omega)+\chi_{j_xj_y}(\omega)\Omega\delta m_y(\omega)$ along the $x$ axis, and similarly along the $y$ axis, where $\chi_{jj}$ is the spin-spin response function at a finite $\Omega$ (that is we are developing a perturbation theory for small variations in the direction of the exchange field, not assuming the smallness of its magnitude $\Omega$). For a system which is spin-rotationally invariant around the $z$ axis, $\chi_{j_xj_x}=\chi_{j_yj_y}$, which will be assumed in the following. The spin density $\mathbf{j}$, in turn, corresponds to the effective field $\mathbf{H}_{\text{eff}}=(\Omega/M_s)\mathbf{j}$ which gives a contribution to the magnetization equation of motion (\ref{llg}). In the low-frequency limit, it translates into the damping coefficient
\begin{equation}
\alpha=(\gamma\Omega^2/M_s)\lim_{\omega\rightarrow0}\mathrm{Im}\chi_{j_xj_x}(\omega)/\omega\,.
\label{a0}
\end{equation}
The susceptibility $\chi_{j_xj_y}$ renormalizes the
local-spin gyromagnetic ratio $\gamma$ if the carrier-spin density is
comparable to the local-spin density. Eq.~(\ref{a0}) can also be
obtained by equating the energy dissipated into the itinerant degrees
of freedom by the moving magnetization and the work done by an rf
magnetic field applied against the viscous Gilbert term in
Eq.~(\ref{llg}), at a steady magnetic precession.
Eq.~(\ref{a0}) is the most basic equation in this
paper and may be taken as the definition of $\alpha$. In the following,
we formally evaluate $\alpha$ for electron and hole systems and
discuss its dependence on the disorder composition.

\begin{figure}
\includegraphics[width=0.9\linewidth,clip=]{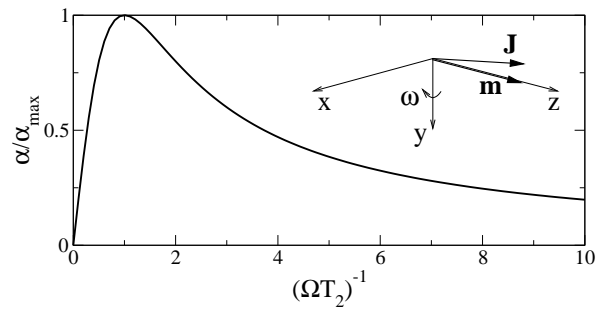}
\caption{\label{fig} Gilbert damping, Eq.~(\ref{a}), in units of $\alpha_{\max}=\gamma j_0/(2M_s)$ as a function of the normalized spin-flip rate. Inset: Geometry of the model.}
\end{figure}

In the absence of spin-orbit interaction in the band structure, $\gamma_2=0$, the average spin density moves in the exchange field $\Omega$ as
\begin{equation}
\partial_t\mathbf{j}=\Omega\mathbf{j}\times\mathbf{m}(t)-\left[\mathbf{j}-j_0\mathbf{m}(t)\right]/T_2\,,
\label{T2}
\end{equation}
where the last term is a phenomenological relaxation due to impurities, characterized by the transverse spin-flip time $T_2$. We assume here that $\mathbf{m}(t)$ undergoes a slow motion (on the scales set by frequencies $\Omega$ and $T_2^{-1}$). It is convenient to transform Eq.~(\ref{T2}) into the frame of reference (for spin variables) moving together with $\mathbf{m}(t)$: If, for example, $\mathbf{m}$ is instantaneously rotating with frequency $\omega$ around the $y$ axis in the laboratory frame, see Fig.~\ref{fig} inset, it is stationary in the rotating frame and there is a new (Larmor) term $\omega\mathbf{j}\times\boldsymbol{\hat{y}}$ on the right-hand side of Eq.~(\ref{T2}), which tries to polarize spins along the $y$ axis. Because the motion is slow, we can solve for $\mathbf{j}$ as the (instantaneous) stationary state in the moving frame of reference. We then find the torque that $\mathbf{j}$ exerts on $\mathbf{m}$ to get
\begin{equation}
\alpha=-\gamma\tilde{\chi}_{j_xj_y}\Omega/M_s=\Omega T_2\left[1+(\Omega T_2)^2\right]^{-1}\gamma j_0/M_s\,,
\label{a}
\end{equation}
where $\tilde{\chi}_{j_xj_y}$ is the stationary (real-valued) response function in the rotating frame, for $\mathbf{m}$ pointing along the $z$ axis. We have thus reduced the calculation of the time-dependent response in the laboratory frame, Eq.~(\ref{a0}), to the static response in the rotating frame. Such a transformation can be done in general for a spin-rotationally--invariant Hamiltonian with no spin-orbit interaction in the band structure. We plot Eq.~(\ref{a}) in Fig.~\ref{fig}. The equilibrium spin density $j_0$ can be calculated from the specific form of the Hamiltonian. $\alpha$ vanishes at both small and large spin-flip rates.

The damping parameter (\ref{a}) scales differently with the spin-flip
rate depending on how it compares with the exchange energy. The low
spin-flip rate regime, $\alpha\propto T_2^{-1}$, is
completely analogous to the ``spin-pumping'' damping\cite{Tserkovnyak:prl021}
of thin films in contact with a
``spin-sink'' conductor: The moving magnetization pumps spins into the
itinerant carriers at a constant rate, which are then relaxed with a
probability $\propto T_2^{-1}$ before exchanging spins with
the ferromagnet. The difference is that now the spins are ``pumped''
into the ferromagnet's own delocalized states. The other limit,
$\alpha\propto T_2$, is simple to understand since
$\mathbf{j}(t)\approx \mathbf{j}_0(t)-T_2
\partial_t\mathbf{j}_0(t)$, in the laboratory frame when the
relaxation rate dominates the dynamics of
$\mathbf{j}(t)$. $\alpha\propto T_2$ then follows from the
torque $\propto(\mathbf{j}-\mathbf{j}_0)\times \mathbf{m}$.
This is analogous to the
``breathing Fermi-surface'' mechanism\cite{Kunes:prb02} of the itinerant
carriers, which try to accommodate the changing magnetization
direction but are lagging behind with a delay time of $T_2$.
In the presence of an anisotropic spin-orbit interaction in the crystal field of a metallic ferromagnet, the breathing Fermi surface gives an additional contribution to damping, which scales linearly with the band-structure relaxation time.\cite{Kunes:prb02}

It is interesting to note that Eq.~(\ref{a}) reduces to the
random-phase approximation result
for the long-wavelength magnon lifetime due to the $s-d$ interaction
with spin-$1/2$ conduction electrons,
which was obtained in Ref.~\onlinecite{Heinrich:pss67} using a fully
quantum-mechanical description:
\begin{equation}
\alpha=\Omega T_2\left[1+(\Omega T_2)^2\right]^{-1}
\gamma\left[\Omega m^\ast k_F/(4\pi^2 \hbar)\right]/M_s\,,
\label{rpa}
\end{equation}
where $m^\ast$ is the band-structure mass and $k_F$ is the Fermi wave
vector, and it was assumed that $\hbar\Omega\ll E_F$ (the Fermi
energy). The quantity in the second square brackets
is just the total carrier spin density.
Eq.~(\ref{rpa}) was used in Ref.~\onlinecite{Ingvarsson:prb02}
to explain the measured damping in thin
permalloy films, which scaled linearly with the film resistivity, as
expected due to the $T_2^{-1}$ prefactor in the relevant
limit of a large exchange energy, $\Omega\gg T_2^{-1}$, in
the transition-metal ferromagnets. Unlike the case of the ferromagnetic
semiconductors, the direct application of the $s-d$ model result is
however questionable in the case of itinerant ferromagnetism in
the transition metals where the separation between the magnetic and
conducting orbitals is unphysical.

We now turn to a discussion of the application of these results to
magnetization relaxation in hole-doped magnetic semiconductor (Ga,Mn)As.
Let us first make a rough estimate of the damping coefficient using
Eq.~(\ref{a}): The largest achievable value of
$\alpha_{\max}=\gamma j_0/(2M_s)$ occurs when
the holes are fully polarized giving $\alpha_{\max}\sim 0.1-0.3$,
roughly one third the ratio of the hole to the substitutional Mn concentrations.
For realistic samples with a spin polarization of the order of unity,
therefore, $\alpha_{\max}\sim 0.1$. The damping $\alpha$ is further
suppressed by the factor
$\alpha/\alpha_{\max}=2\Omega T_2[1+(\Omega T_2)^2]^{-1}<1$.
For clean bulk samples of GaAs, the spin-flip relaxation time
is $\sim 100$~fs.\cite{Hilton:prl03}
For approximately 5~\% Mn doping,
$\hbar\Omega\sim0.1$~eV,\cite{Dietl:prb01}
so that $\Omega T_2\sim10$,
puts one in the $\alpha \propto T_2^{-1}$
regime with $\alpha\sim0.01$.
Shorter spin-flip times would thus result in larger damping.
Experimentally, the impurity scattering is likely to be
the easiest parameter to vary in order to ``engineer'' a desired
$\alpha$.  For a bulk sample, the strong spin-orbit coupling $\gamma_2$,
however, makes the validity of the phenomenological equation (\ref{T2}),
and thus result (\ref{a}), questionable. Besides, the crystal anisotropy
would require a further refinement of the analysis. We thus have to return
to Eq.~(\ref{a0}) in order to derive a reliable result.
Evaluating the response function for a noninteracting Hamiltonian yields
\begin{eqnarray}
\alpha&=&\frac{\gamma\Omega^2}{M_sV}\lim_{\omega\rightarrow0}\frac{\pi}{\omega}\sum_{ij}\left|\langle i|J_x|j\rangle\right|^2\left[f(\epsilon_i)-f(\epsilon_j)\right]\nonumber\\&&\times\delta(\hbar\omega+\epsilon_i-\epsilon_j)\,,
\label{aij}
\end{eqnarray}
where $f(\epsilon)$ is the Fermi-Dirac distribution and $i$, $j$ label one-particle eigenstates in the sample of volume $V$. If the lattice vector $\mathbf{k}$ is conserved, $\sum_{ij}/V=\sum_{ab}\int d^3k/(2\pi)^3$, where $a$, $b$ label spin states. For a perfect crystal, therefore, $\alpha$ vanishes, as expected (unless there is a finite-measure Fermi surface area with a spin degeneracy). The role of the relaxation on lattice defects was formally introduced in Ref.~\onlinecite{Sinova:prb04} by broadened one-particle spectral functions, $A_{\mathbf{k}a}(\epsilon)=\Gamma/[(\epsilon-\epsilon_{\mathbf{k}a})^2+\Gamma^2/4]$, as follows:
\begin{eqnarray}
\alpha&=&\frac{\gamma\Omega^2}{M_s}\lim_{\omega\rightarrow0}\frac{\pi}{\omega}\int\frac{d^3k}{(2\pi)^3}\sum_{ab}\left|\langle \mathbf{k}a|J_x|\mathbf{k}b\rangle\right|^2\int\frac{d\epsilon}{(2\pi)^2}\nonumber\\&&\times A_{\mathbf{k}a}(\epsilon)A_{\mathbf{k}b}(\epsilon+\hbar\omega)\left[f(\epsilon)-f(\epsilon+\hbar\omega)\right]\,,
\label{aG}
\end{eqnarray}
which was obtained by evaluating the local-spin susceptibility after integrating out the itinerant-carrier degrees of freedom. They\cite{Sinova:prb04} find a nonmonotonic behavior of $\alpha$ as a function of the phenomenological scattering rate $\Gamma$ for a realistic (Ga,Mn)As band structure: $\alpha\propto\Gamma^{-1}$ as $\Gamma\rightarrow0$ (after taking the $\omega\rightarrow0$ limit first) and, after passing through a minimum, $\alpha$ increases monotonically with large $\Gamma$. Eq.~(\ref{aG}) however appears to have a problem for the large momentum-scattering rate, $\Gamma$, asymptotic: When $\Gamma\gg\gamma_2$, the D'yakonov-Perel\cite{Dyakonov:jetp71} spin-relaxation rate for Hamiltonian (\ref{H0}) scales as $\gamma_2^2/\Gamma$; the spin-spin response corresponding to Eq.~(\ref{aG}), on the other hand, has a $1/\Gamma$ cutoff in the time domain, resulting in a spin-relaxation rate growing linearly with $\Gamma$.

In our discussion, we have assumed that the ferromagnetic
magnetization is moving without specifying the exact mechanism of how
the motion is initiated. Doing this by, e.g., applying an external
magnetic field (with a large dc and small rf components)
will of course affect the form of the Hamiltonian
(\ref{H}) for the itinerant carriers. Our results for the Gilbert
damping will stay unaffected, however, as long as the exchange energy
$\hbar\Omega$ is much larger than the
carrier Zeeman splitting in the applied field,
and the ferromagnetic magnetization is mostly supplied by the localized
orbitals (otherwise one has to take into account the energy pumped by the rf
field into the carrier-magnetization dynamics).

We are grateful to G.~E.~W.~Bauer, A.~Brataas, and B.~Heinrich for
stimulating discussions. This work is supported in part by
the Harvard Society of Fellows, the DARPA award No. MDA 972-01-1-0024,
and NSF Grants PHY 01-17795 and DMR 02-33773.

\end{document}